\documentclass[conference,letterpaper]{IEEEtran}
\usepackage[font=small,skip=10pt]{caption}
\usepackage{enumitem}
\usepackage{color}
\usepackage{enumitem}
\usepackage{amsmath}
\usepackage{graphicx}

\usepackage{algorithm}
\usepackage{algorithmic}

\newcommand{\comment}[1]{}

\begin{document}

%
\title{Approximation Algorithm for\\ N-distance Minimal Vertex Cover Problem}

\author{
\IEEEauthorblockN{Tarun Yadav}
\IEEEauthorblockA{Scientist, Scientific Analysis Group\\Defence R \& D Organisation, INDIA\\Email: tarunyadav@sag.drdo.in}

\and

\IEEEauthorblockN{Koustav Sadhukhan\IEEEauthorrefmark{1}, Rao Arvind Mallari\IEEEauthorrefmark{2}}
\IEEEauthorblockA{Scientist, Defence Research and\\Development Organisation, INDIA\\Email: \IEEEauthorrefmark{1}koustavsadhukhan@hqr.drdo.in, \IEEEauthorrefmark{2}arvindrao@hqr.drdo.in}}

\maketitle

\begin{abstract}
Evolution of large scale networks demand for efficient way of communication in the networks. One way to propagate information in the network is to find vertex cover. In this paper we describe a variant of vertex cover problem naming it \textit{N-distance Vertex Minimal Cover(N-MVC) Problem} to optimize information propagation throughout the network.  A minimum subset of vertices of a unweighted and undirected graph $G=(V,E)$ is called \textit{N-MVC} if $\forall v \in V$,  $v$ is at distance $\leq N$ from at least one of the the vertices in \textit{N-MVC}.  In the following paper, this problem is defined, formulated and an approximation algorithm is proposed with discussion on its correctness and upper bound. 

\end{abstract}

\IEEEpeerreviewmaketitle

\section{Introduction}

Network such as Internet, human body, malicious botnet, mobile networks are part of everyday life. Networks in physical world are mapped to graphs in computer world. A graph $G (V,E)$  is set of nodes called vertices($V$) and connections between these nodes called Edges($E$).
Essence of network is communication between nodes, in case of large networks it is a big challenge to perform this task efficiently. Primarily there are two objectives, first information should reach to each node of graph and second is to perform this task efficiently with minimum resources and given constraints. Given unlimited resources and no constraints, information can propagate to each node but this is not the case in real life scenarios. In practice, there is need to select some of the nodes which can propagate given information to other nodes. Example of such scenarios can be found in social networks\cite{7}, P2P botnets and similar densely connected graphs of various networks. As a specific example, consider the case where one has to select the minimal set of influential nodes in a social network such that some critical information is propagated to all nodes in the network in a finite number of hops. 

One solution of this problem is to determine an approximate Minimal Vertex Cover(\textit{MVC}) of the network and use the nodes in MVC for propagating information. Due to the property of \textit{MVC}, the information will be propagated to all nodes in the network in a single hop. But in practice, the cardinality of \textit{MVC} in huge networks will be very large. Hence, the resources used to propagate information using a single hop from vertices of \textit{MVC} will be very high. A smaller set of nodes can be used to propagate the information in multiple hops, to reduce the number of resources used. The challenge is to find a set of nodes given a constraint of the maximum hops($N$), before which information has to be propagated to all nodes in the network. We propose  a solution to estimate this smaller set of nodes and term  it as approximated \textit{N-distance minimum vertex cover (N-MVC)}, where $N$ is the maximum  number of hops within which information has to reach all nodes in the network. If the network is static and don't change over time then once computed nodes in \textit{N-MVC} require propagation capability, all other node may behave as sink nodes (don't propagate the information) whereas in dynamic networks where nodes or connections are changed over time, \textit{N-MVC} needs to be  recomputed as soon as new any change happens. We can say $N$ is capacity of nodes to propagate the information. We are considering homogeneous situation where $N$ is same for all nodes. In this paper, we discuss the problem statement, present the approximation algorithm, correctness and discuss upper bound on the solution. 			 

One of the similar type of problem is \textit{k-path Vertex Cover}\cite{2} and discussed\cite{4} \cite{5} thoroughly. \textit{k-path Vertex Cover} ensures existence of special(having some defined properties) nodes in any path of k vertices in the network.  Variants of Vertex cover problem are part of research in domains related to secure communication in sensor networks\cite{6}, topology analysis of malicious bots network and security and resources optimization in various types of network.    

This paper is organized into 4 sections. Section II defines the problem statement, Section III describe the algorithm in detail, Section IV proves correctness of the algorithm using contradiction, Section V discusses the upper bound for the solution with reference to N . The paper ends with summary and concluding remarks in Section VI.

\section{N-distance Minimal Vertex Cover Problem }
Given an unweighted and undirected graph $G(V,E)$, we define the following.

\textbf{Vertex Cover (VC):} The Vertex cover of Graph $G(V,E)$ is a subset of vertices $S \subseteq V(G)$ such that every edge has atleast one endpoint in $S$, that is  $(u,v) \in E(G) \implies  u \in S\ \lor\  v \in S  $. 
Alternatively,  vertex cover of a given Graph $G(V,E)$ is a set S of vertices such that any vertex of the graph either $ \in S$ or at most $1$ hop distance(or edge) away from at least one of the vertices in $S$. 

\textbf{Minimum Vertex Cover Problem(MVC)\cite{1}:} For a graph $G$, Minimum Vertex Cover Problem(MVCP) is the optimization problem of finding the vertex cover of $G$ with the least possible cardinality.

\textbf{ N-distance  Vertex Cover(N-VC):} N-distance Vertex Cover(N-VC) for a given Graph $G(V,E)$ is a subset of vertices $S \subseteq V(G)$ such that every vertex is either in $S$ or at most at a distance of $N$ away from at least one of the vertices in vertex cover.\\
$\forall v \in V(G)$, either $v \in S$ or $\exists u \in S$ such that $d(v,u) \leq N$   where $d(u,v)$ denotes the geodesic distance between vertices u and v. 

\textbf{ N-distance  Minimum Vertex Cover Problem(N-MVC):} N-distance Minimum Vertex Cover(N-MVC) Problem for a graph G is the optimization problem of finding the N-distance Vertex cover with the least possible cardinality.

\textbf{ Explanation of Problem Statement:}\textit{ N-distance Minimal Vertex Cover (N-MVC)} is a set of minimum nodes($S$), such that every node of network either belongs to \textit{N-MVC} or is at a maximum of $N$ hops away from atleast one node in $S$.  In other words it can be said that in any arbitrary network, information($I$) propagated to $N$ nodes in  \textit{N-MVC} will guarantee that each and every node in the network will possess the information(I). This is the problem statement, that the authors would like to solve in this paper.

The problem being formulated and solved is specifically for unweighted and undirected graphs. Distance between two nodes $u,v$ in a graph $G$ is denoted as $d(u,v)$ and defined as the number of edges on the geodesic (shortest path), if it exists, connecting them. If no geodesic exists, as per convention distance is taken to be infinite. Since we are dealing with undirected graphs, $d(u,v)=d(v,u)$.

\begin{figure}
\includegraphics[scale=.5]{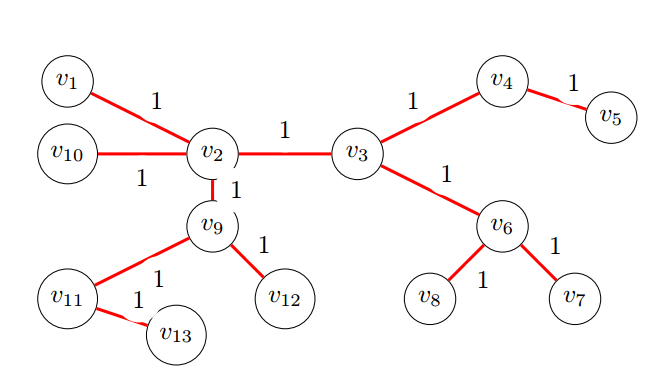}
%
 \caption{An example graph to understand N-distance vertex cover} \label{example}
\vspace{-.6cm}
\end{figure}

\textbf{Example:}
In example graph Fig. \ref{example} 3-distance minimum vertex cover is  $S=\{v_4,v_9\}$. It is easily verifiable that $\forall v \in V(G), \exists u \in S$ such that $d(u,v) \leq 3$.

Similarly, the reader can verify that in the same graph 2-distance minimum vertex cover is $S=\{v_3,v_9\}$.

In next section, we present an approximation algorithm for \textit{N-MVCP} because \textit{N-MVCP} is a \textit{NP-Complete} problem. As we will discuss a reduction from \textit{N-distance} to \textit{1-distance} \textit{MVCP}, which is original \textit{MVCP}, it is implied that \textit{N-MVCP} is \textit{NP-Complete}.  

\section{Approximation Algorithm to find N-distance Minimal Vertex Cover }
Before discussing the approximation algorithm, we will revisit definitions and concepts of graph theory being used.
\\
\textbf{Walk}: Given a graph $G=(V,E)$, a walk $W(v_0,v_n)$ joining $v_0$ and $v_n$ is defined as an alternating sequence of vertices and edges of G\\
$W(v_0,v_n)=v_0,e_1,v_1,e_2,v_2,\cdots,e_n,v_n$ \\
such that $e_i=(v_{i-1},v_i),1\leq i \leq n$. The length of a walk $W$ denoted by $\ell(W)$ is the number of edges in W.
\\
\textbf{Trail:}A walk $W(v_0,v_n)$ is called a trail if all edges in the walk are different.\\
\textbf{N-Trail:}A walk $W(v_0,v_N)$ is called a \textit{N-Trail} if there are N in the walk and all are different.\\ 
\textbf{Degree($v,E$):} Number of edges of $E$ incident to the vertex $v$. In this paper Degree($v,E$) is denoted as $deg(v,E)$.

For a given $G(V,E)$ and value of $N$ ( where G has at least one  $N-path$ as described in Algorithm \ref{Npath}  and $N\geq2$  because for $N=1$ algorithm reduces to original vertex cover problem), the approximation algorithm is described as Algorithm \ref{mainalgo} in subsection \ref{algo} which calls Algorithm \ref{Npath} (\ref{Nedges}) and Algorithm \ref{reduction} (\ref{GRA}) as subalgorithms.

\subsection{Finding $N-Trail$ (Trail of length $N$) }
\label{Nedges}

\begin{algorithm}
\caption{Algorithm to Find $N-Trail$ \label{Npath}}
\begin{algorithmic} [1]
\REQUIRE  $G(V,E)$ where $V = \{v_1,v_2,...\}$ , $E = \{e_1,e_2,...\}$
\ENSURE  Trail $E_N$  such that $\ell(E_n)=N$
\STATE $E_N \leftarrow  \phi $
\STATE $E_N \leftarrow E_N \cup e_i$ where  $e_i=\{v_j,v_k\}$  s.t. $e_i \in E$ \& $ deg(v_k,E ) \geq 2$
\REPEAT
\STATE $E^{'}  = \phi $
\STATE   $E^{'} \leftarrow  \{e_i, \ s.t.\  e_i= (v_k, v_m)\  \&\  e_i \in E\}$ 
\STATE   $E_N \leftarrow E_N \cup e_j$ where   $e_j=\{v_k,v_m\} $   s.t. $e_j \in E^{'} $ \& $deg(v_m,E ) \geq 2 $ if $|E_N| < N$     
\STATE   $v_k \gets v_m $
\UNTIL{$|E_N| \neq N$}
\RETURN $E_N$

\end{algorithmic}
\end{algorithm}

\begin{enumerate}
\item{Compute degree of each vertex $v_i \in V$ and store it.}
\item{Randomly pick an edge $e_i=\{v_j,v_k\}$ s.t. degree of $v_k$ is at least 2. Add $e_i$ to $E_N$.}
\item{Pick the vertex $v_k$ which has degree at least 2. Find all edges $E^{'}$ s.t. each edge in $E^{'}$ has $v_k$ as one end point. }
\item{As we know one end point of $e_j$ is $v_k$ and let us say another end point is $v_m$. 
Now pick any edge $e_j \in E^{'} $ s.t. $degree(v_m) \geq 2$. and add it to $E_N$. Now repeat step 3 with $\ v_k \gets v_m$ until size of $E_N \neq N$} 
\item{We define endpoints of $E_N$ as vertices $\{v_x,v_y\}\ s.t.\ deg(v_x)\ and\ deg(v_y) = 1 $ considering only edges $e_i \in E_N$ }
\end{enumerate}

\subsection{Graph Reduction Algorithm}
\label{GRA}

\begin{algorithm}
\caption{Algorithm for Graph Reduction \label{reduction}}
\begin{algorithmic} [1]
\REQUIRE  $G(V,E)$ and $distance(e) =1\; \forall e \in E$
\ENSURE    Reduced Graph $G^{'}(V,E^{*})$

\STATE  $G^{'}(V,E^{*}) \gets G(V,E)$
\STATE loop:
\WHILE{$N \geq 2$}
\IF  { $N-Trail$ not exists in $G^{'}$}
\STATE  $N \leftarrow N-1$
\STATE  go to loop	
\ENDIF
\STATE   $V^{'} \leftarrow \phi$, $V^{''} \leftarrow \phi$, $E^{'} \leftarrow \phi$, $E^{''} \leftarrow \phi$,$E^{'''} \leftarrow \phi$, $E^{''''} \leftarrow \phi$
\STATE   Pick a $N-Trail$ $E_N$ from Algorithm \ref{Npath} with $endpoints(E_N) = (v_0,v_N)$ \& \ $ edges(E_N) \in E$
\STATE   $V^{'} \leftarrow  \{v_i\}\ s.t.\ either\ d(v_0,v_i) = N\  or\  (2 < d(v_0,v_i)  <  N\  \&\  deg(v_i,E)=1)\ \forall v_i \in V$
\STATE   $V^{''} \leftarrow  \{v_j\}\ s.t.\ either\ d(v_j,v_N) = N\  or\  (2 < d(v_j,v_N)  <  N\  \&\  deg(v_j,E)=1) \ \forall v_j \in V$
\STATE   $E^{'} \leftarrow  \{ e = (v_0,v_i) \}\  \forall v_i \in V^{'}$ 
\STATE   $E^{''} \leftarrow \{ e = (v_j,v_N) \}\  \forall v_j \in V^{''}$ 
\STATE   $E^{'''} \leftarrow$ \{$e_i,\  e_i$ are edges of $k-Trail$ from  $v_0$  to $v_i \  \forall   v_i \in V^{'}, e_i \in E$ \& $k \leq N$ \}
\STATE   $E^{''''} \leftarrow$ \{$e_j,\  e_j$ are edges of $k-Trail$ from  $v_j$  to $v_N\  \forall  v_j \in V^{''}$, $e_j \in E \setminus E^{'''}$ \& $k \leq N$ \}
\STATE   $G^{'}(V,E^{*}) \gets G^{'}(V,E^{*} \cup E^{'} \cup E^{''} \setminus E^{'''} \setminus  E^{''''} ) $
\ENDWHILE
\RETURN $G^{'}(V,E^{*})$
\end{algorithmic}
\end{algorithm}

\textbf{Initialization:} 
Initialize $G^{'} \gets G$ and Compute degree of each vertex of $G^{'}$ and store it. Assign each edge of $G^{'}(V,E)$ 1 unit of distance. \\
\begin{enumerate}
\item{Pick a $Trail$ of $N$ connected edges (using Algorithms discussed in subsection \ref{Nedges}) $\{e_1,e_2,...,e_{N}\}$ connecting $N+1$ vertices $\{v_0,v_2,...,v_N\}$ from $G^{'} \, s.t.\ e_i = \{v_i,v_{i+1}\}$ and we define endpoints of $N$ connected edges as $\{v_0,v_N\}$ 
}
\item{Find set of vertices $V^{'}$ and $V^{''}$, which are either $N$ (where $N\geq2$) distance away from ($v_0$ or $v_N$)  or $<N$ distance away  from ($v_0$ or $v_N$) and having degree one, s.t. each vertex from $V^{'}$ or $V^{''}$ is connected to $v_1$ or $v_{N+1}$  respectively by a trail. }
\item{For each vertex $v_i$ from $V^{'}$, add an edge between $v_0$ and $v_i$ and mark edges connecting $v_0$ and $v_i$ for deletion and For each vertex $v_j$ from $V^{''}$ add an edge between $v_j$ and $v_N$ and mark edges connecting $v_j$ and $v_N$ for deletion. Now delete all the edges marked for deletion and recompute the degree and update the degree table w.r.t edges with distance assigned ($e\in E$).} 
\item{Repeat step 1,2,3 with new $Trail$ of length $N$  using algorithm \ref{Npath} s.t. $edges(Trail) \in E $.\\
\textbf{Note:} We will not use newly added edges(not assigned any distance) of G to form $N-Trail$ in Algorithm \ref{Npath}}
\item{if no \textit{N-Trail} is found then go to step 1 with $N \gets N-1 \  (\ N\geq2)$}

\end{enumerate}

\subsection{Approximation algorithm for N-distance Minimal Vertex Cover Problem (NMVCP)} 
\label{algo}
Approximation algorithm is application of graph reduction and then Approx-Vertex-Cover\cite{3} algorithm subsequently.

\begin{algorithm}
\caption{Approximation Algorithm for NMVCP \label{mainalgo}}
\begin{algorithmic} [1]
\REQUIRE  $G(V,E)$ s.t. $V = \{v_1,v_2,...\}$ , $E = \{e_1,e_2,...\}$
\ENSURE    Approximated Solution of \textit{NMVCP} for $G(V,E)$

\STATE   $G^{'}(V,E^{'}) \gets$  Graph Reduction Algorithm \ref{reduction} with input $G(V,E)$
\STATE   \textit{NMVC} $\gets$ \textit{Approx-Vertex-Cover}\cite{3} Algorithm \ref{approx} with input  $G^{'}(V,E^{'}$)
\RETURN NMVC
\end{algorithmic}
\end{algorithm}

\begin{algorithm}
\caption{Approx-Vertex-Cover(G) \cite{3} \label{approx}}
\begin{algorithmic} [1]
\REQUIRE  $G(V,E)$
\ENSURE   Approx-Vertex-Cover solution $AVC$

\STATE $AVC \gets \phi$
\WHILE {$E \neq \phi$}
\STATE pick any $(u, v) \in E$
\STATE $AVC \gets AVC \cup \{u, v\}$
\STATE delete all edges incident to either u or v
\ENDWHILE
\RETURN AVC
\end{algorithmic}
\end{algorithm}


\subsection{Example}
\label{test}
\begin{figure}
\includegraphics[scale=.5]{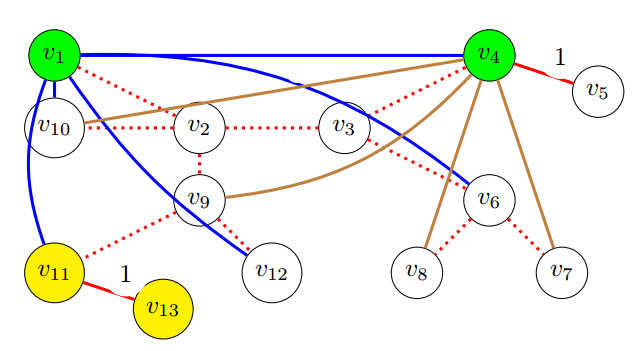}
 \caption{Graph after one iteration of Graph Reduction Algorithm} \label{example_1}
\vspace{-.6cm}
\end{figure}

\begin{figure}
\includegraphics[scale=.5]{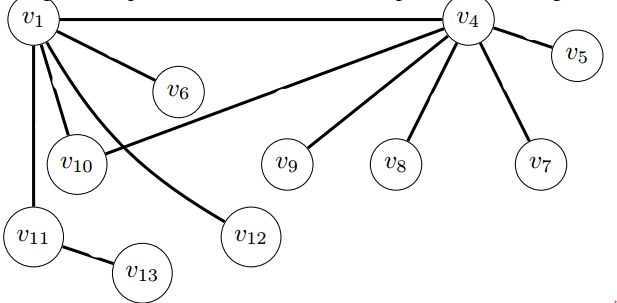}
%
%
%
%
%
 \caption{Graph after one iteration of Graph Reduction Algorithm} \label{example_2}
\vspace{-.6cm}
\end{figure}

As algorithm described in subsection \ref{algo} (Algorithm \ref{mainalgo}) to find \textit{N-distance vertex cover}, first step is to apply graph reduction algorithm to the given graph. If we consider the graph shown in Fig. \ref{example} as input then step 1 of algorithm \ref{mainalgo} for graph reduction (algorithm \ref{reduction}) with $N=3$ will proceed as follows:\\
\textbf{Initialization:} $G^{'} \gets G$ and each edge is assigned a distance of 1 unit and compute degree of each vertex.
\begin{enumerate}
\item{
From algorithm \ref{Npath} we  get  \textit{N(=3)-path} as described in subsection \ref{Nedges}. Suppose we get set of three edges $\{ e_{v_1-v_2}, e_{v_2-v_3},e_{v_3-v_4}\}$ as one of the \textit{N-Trail}. End point of this \textit{N-Trail} are $\{v_1,v_4 \}$}
\item{
\begin{enumerate}
\item{Vertices at $N$ (=3) distance away from $v_1$ are $\{v_4,v_6,v_{11},v_{12}\}$\\
Vertices at $<N$ ($=3$) but $\geq2$ distance from $v_1$ and with degree 1 are $\{v_{10}\}$}	
\\Therefore $V^{'}\ =\ \{v_4,v_6,v_{10},v_{11},v_{12}\}$
\item{Vertices at $N$ ($=3$) distance away from $v_4$ are $\{v_1,v_7,v_8,v_9,v_{10}\}$ \\
Vertices at $<N$ ($=3$) $\geq2$ distance from $v_4$ and with degree 1 are $\{\phi\}$}	
\\Therefore $V^{''}\ =\ \{v_1,v_7,v_8,v_9,v_{10}\}$
\end{enumerate}
   }
\item{
\begin{enumerate}
\item {Add an edge $e_{v_1-v_i} \forall v_i \in V^{'}$  and mark connecting edges (of corresponding Trail) for deletion and add an edge $e_{v_j-v_4} \forall v_j \in V^{''}$ and mark connecting edges for deletion as shown in Fig. \ref{example_1}\\
$E^{'} \leftarrow \{ e_{v_1-v_4},  e_{v_1-v_6},   e_{v_1-v_{10}}, e_{v_1-v_{11}}, e_{v_1-v_{12}}\}$\\
$E^{"} \leftarrow \{ e_{v_4- v_1},  e_{v_4-v_7},  e_{v_4-v_8},  e_{v_4-v_9},  e_{v_4-v_{10}}\}$\\
$E^{'''} \leftarrow \{ e_{v_1-v_2}, e_{v_2-v_3}, e_{v_3-v_4}, e_{v_3-v_6}, e_{v_2-v_9}, e_{v_9-v_{12}},\\ e_{v_9-v_{11}}, e_{v_2-v_{10}}\}$\\
$E^{''''} \leftarrow \{ e_{v_6-v_7},  e_{v_6-v_8}\}$\\
}
\item{After adding the edges ($E^{'}\ \&\ E^{"} $), remove the edges ($E^{'''}\ \&\ E^{''''} $) marked for deletion and update the degree table considering  edges  from $G$  ($e \in E$).}
\end{enumerate}
}
\item{Now if we search for a new \textit{N(=3)-Trail} in reduced graph in Fig. \ref{example_2}, there are no such edges.  There is no  \textit{N(=2)-path} also, therefore only edges left are $e_{v_4-v_5}$ and $e_{v_{11}-v_{13}}$ with $N(=1)$ which is not considered in algorithm.  } 
\end{enumerate}

%

After graph reduction process $G^{'}$ is the graph as shown in Fig. \ref{example_2} but without distances assigned to edges. Now as described in step 2 of algorithm \ref{mainalgo} Approx-Vertex-Cover( algorithm \ref{approx}) for $G^{'}$ will return the solution \textit{NMVC}.  \\

\textit{Approx-Vertex-Cover Algorithm on $G^{'}$} \\
\begin{enumerate}
\item{Select an edge randomly from $G^{'}$ and add its endpoints to the solution \textit{AVC} and remove all edges connected to  the endpoints of this edge. Lets pick the edge $e_{v_1-v_4}$ then all edged connected to $v_1$ or $v_4$ will be removed and we can see only one edge will remain after this process $e_{v_{11}-v_{13}}$ }
\item{Now we need to pick another edge randomly but as we know only one edge is there in the graph therefore we need to pick $e_{v_{11}-v_{13}}$ and add its endpoints to the solution \textit{AVC}}
\item{Now no edges is remaining in graph therefore solution \textit{AVC}= $\{v_1,v_4,v_{11},v_{13}\}$ which is \textit{NMVC} in algorithm \ref{mainalgo}}
\end{enumerate} 
Here we can see optimal solution is for \textit{NMVC} is $ \{v_1,v_4,v_{11}\}$ but  approx-vertex-cover algorithm outputs approximate solution with 1 extra vertex.

\section{Proof of Correctness of Algorithm}
We will prove the correctness of proposed algorithm using contradiction.\\
By graph reduction described in algorithm \ref{reduction} we are reducing $G(V,E) \to G^{'}(V,E^{'})$ and solving $G^{'}$ for Vertex Cover using \textit{Approx-Vertex-Cover} algorithm which will provide solution for \textit{N-distance vertex cover} for $G$.\\
By reduction algorithm any edge $e^{'}=\{u,v\} \in E^{'}$  is either an edge from original graph ($e \in E$) or added by graph reduction algorithm.\\
Lets assume there is an edge $e \in E$ which is not covered by \textit{N-distance vertex cover solution(NMVC)} given by the proposed algorithm. It means both the endpoint of $e$ are $> N$ distance from each vertex in \textit{NMVC}.\\
From reduction algorithm we can say  either $e \in E^{'}$ or $e$ is removed during reduction $G \to G^{'}$\\
if $e \in E^{'}$ then e has to be covered by approx-vertex-cover algorithm for $G^{'}(V,E^{'})$ because of correctness of \textit{Approx-Vertex-Cover} algorithm(contradiction to assumption).\\
If $e$ is removed during graph reduction algorithm then by properties of reduction algorithm\\
\begin{enumerate}
\item{An edge is only removed when it's endpoints are $\leq N$ distance from one of the the endpoints of $E_N$.}
\item{During reduction all the edges of $E_N$ are removed but one new edge is added between endpoints of $E_N$}\\
\end{enumerate}
 From subsection \ref{GRA} step 3, $e$ could be an edge connecting $v_0$ and $v_i$ or connecting $v_j$ and $v_N$. When edges connecting these vertices will be removed, one edge connecting endpoint vertices will be added. So, one of these endpoints (of newly added edge) has to be in solution (by Approx-Vertex-Cover Algorithm property) and $e$ is $\leq N$ distance from both endpoints. Therefore $e$ is $\leq N$ distance from one of the vertex in solution which is contradiction to the assumption. \\

\section{Discussion on Upper bound on Solution}

\begin{figure}
\includegraphics[scale=.5]{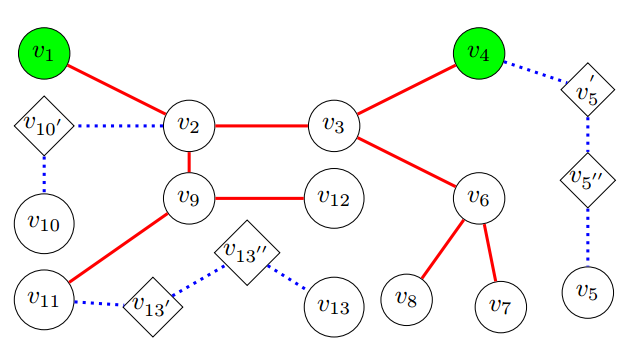}
 \caption{Graph $G^{''}$ - tretched version of $G$ w.r.t example descrined in subection \ref{test} } \label{stretch}
\vspace{-.6cm}
\end{figure}

Given graph instance($I$) $G(V,E)$, solution for \textit{N-distance vertex cover} depends on reduction $G \to G^{'}$. From \textit{Approx-Vertex-Cover} upper bound on solution for $G(V,E):$\\
\begin{equation}  \label{eqn1}
OPT(I) \geq |M|
\end{equation}
\begin{equation} \label{eqn2}
A(I) = 2|M| \leq 2\ OPT(I)
\end{equation}
where $M$ is maximal matching for $G$ and $|M|$ is size of maximal matching. $OPT(I)$ is the optimal solution for the given instarnace $I.$\\
After Graph Reduction Algorithm performed $G \to G^{'}$, new instance ($I^{'}$) is  $G^{'}(V,E^{'}).$
To get final solution, \textit{Approx-Vertex-Cover} algorithm is applied on $G^{'}$ in algorithm \ref{mainalgo}. Therefore we can say 
\begin{equation}
OPT(I^{'}) \geq |M^{'}|
\end{equation}
\begin{equation}\label{eqn4}
A(I^{'}) = 2|M^{'}| \leq 2\ OPT(I^{'})
\end{equation}
where $M^{'}$ is maximal matching for $G^{'}$ and $|M^{'}|$ is size of maximal matching. $OPT(I^{'})$ is the optimal solution for the given instarnace $I^{'}.$\\
We know each edge of maximal matching in $M^{'}$ is added by removing a \textit{k-Trail\ ($k\leq N$)} of $\leq N$ connected edges in $G$ selected using algorithm \ref{Npath}. \\
We will modify graph $G \to G^{''}$ s.t. \textit{N-distance vertex cover} solutions for $G^{'}$ and  $G^{''}$ are same. We initialize $G^{''} \gets G$\\
As we know when graph reduction algorithm executes 2 conditions are checked when an edge is added between vertices with distance $<N:$\\
\begin{enumerate}
\item{\textit{Algorithm \ref{reduction} steps 10 \& 11:} Vertices with distance $<N$ but $degree=1$ are added to $V^{'}$ and $V^{''}$ respectively and  then edges $E^{'}$ and $E^{''}$ are added in step 12 \& 13.}
\item{\textit{Algorithm \ref{reduction} step 4:} if there is no \textit{N-path}, $N$ is decreased (to $n$)  and algorithm is repeated. When $N$ is decreased and reduction algorithm is processed, new edges are added between the vertices with distance $< N$(= $n$).} 
\end{enumerate}
While executing algorithm in both the conditions, we will extend the vertex other than endpoint of $E_N$ in $G^{''}$ by splitting the edge connected to that endpoint into $ N-n$ edges and vertices as shown in Fig. \ref{stretch}. This extension also apply to the vertices  at $1$ unit distance($N=1$) which we have avoided in reduction algorithm.  By this construction we have made exactly \textit{N-Trail} to one of the endpoint of $E_N$ to each vertex in $V^{'}$ \& $V^{''}$. \\
The above construction guarantees that in $G^{'}$ every maximal matching is an edge added by removal of exactly $N$  edges of a \textit{N-Trail} in $G^{''}$. 
If we apply \textit{Approx-Vertex-Cover} algorithm to $G^{''}$ then 
\begin{equation}
OPT(I^{''}) \geq |M^{''}|
\end{equation}
\begin{equation} \label{eqn6}
A(I^{''}) = 2|M^{''}| \leq 2\ OPT(I^{''})
\end{equation}
where $M^{''}$ is maximal matching for $G^{''}$ and $|M^{''}|$ is size of maximal matching.  $OPT(I^{''})$ is the optimal solution for the given instance $I^{''}.$\\
As we discussed each edge of maximal matching in $M^{''}$ is replacement of a \textit{N-Trail} in $G^{''}$. Therefore, from the construction of $G^{''}$ we can write
\begin{equation} 
|M''| \geq |M'|\  \ast\  N/2\ \  
\end{equation}
\begin{equation} \label{eqn8}
|M'| \leq \ 2/N  \ast\  |M''|
\end{equation}
 ( $N\gets N+1$  if  N is odd )\\
from equations \ref{eqn4} and \ref{eqn8}
\begin{equation}\label{eqn9}
A(I^{'}) = 2|M^{'}| \leq 2 \ast (2/N)\ast|M''|  
\end{equation}
From equations \ref{eqn6} and \ref{eqn9}
\begin{equation}\label{eqn10}
A(I^{'}) \leq (2/N)\ast2|M''|\  \leq (4/N)OPT(I^{''})  
\end{equation}
from equation \ref{eqn10} we can say for a given graph $G$ if we extend $G$ to $G^{''}$ and reduce $G$ to $G^{'}$ then solution to \textit{N-distance Vertex Cover Problem} for $G$ is the solution from \textit{Approx-Vertex-Cover} algorithm for $G^{'}$ which have upper bound w.r.t $G^{''}$, which is extended version of $G$.   \\

\textit{\textbf{Utility of Extended Graph:}} We constructed extended graph to discuss upper bound on solution but there are utilities of extended version of a graph. As we discussed extending a graph ensures the traversal of exact \textit{N-Trail} which inherently suggest a process to maximize a network's capability to propagate the information. Fig. \ref{example_2} shows the extended graph where dotted edges are extended edges and diamond shaped nodes are extended nodes. In a network scenario if more no. of nodes are to be added then these extended nodes are the possible places where one should add the new nodes without worrying about propagating the information to the new nodes. These new nodes could be sink nodes which don't require capability to propagate information.
\section{Summary} 
This paper presents a variant of vertex cover problem called \textit{N-distance vertex cover problem} which addresses challenges in networks to propagate information efficiently. We proposed a solution by approximation algorithm using graph reduction and \textit{Approx-Vertex-Cover} algorithm. Correctness of proposed solution is proved using contradiction and upper bound is also discussed by construction of extended graph.

\end{document}